# Supercell periodicity biases octahedral tilting and energetics in high-entropy perovskite oxides

Hanzo Tsubota[1], Wai Kian Tan[2,3], Hiroyuki Muto[1,2,3], Atsunori Matsuda[1,2], Tom Ichibha[4], Kenta Hongo[5], Ryo Maezono[6], Liangzhi Kou[7], Go Kawamura[1,2*]

[1] Department of Electrical and Electronic Information Engineering, Toyohashi University of Technology, Toyohashi, Aichi 441-8580, Japan

[2] Institute for Research on Next Generation Semiconductor and Sensing Science, Toyohashi University of Technology, Toyohashi, Aichi 441-8580, Japan

[3] Institute of Liberal Arts and Sciences, Toyohashi University of Technology, Toyohashi, Aichi 441-8580, Japan

[4] School of Information Science, Japan Advanced Institute of Science and Technology, Nomi, Ishikawa 923-1292, Japan

[5] Center for Advanced Scientific Computing, Japan Advanced Institute of Science and Technology, Nomi, Ishikawa 923-1292, Japan

[6] Graduate Major in Materials and Information Sciences, School of Materials and Chemical Technology, Institute of Science Tokyo, Meguro-ku, Tokyo 152-8550, Japan

[7] School of Mechanical, Medical and Process Engineering, Queensland University of Technology, Brisbane, QLD 4001, Australia

*Corresponding author: Go Kawamura (kawamura.go.km@tut.jp)

## Abstract

Odd-period supercells such as 3 × 3 × 5 are often used for equimolar five-component compositions in computational studies of high-entropy perovskite oxides (HEPOs). However, such cells are known to constrain octahedral tilting because they cannot accommodate all standard tilt patterns (Glazer tilt systems). How strongly this constraint affects HEPOs, whose octahedra respond to local cation environments, remains unclear. Here we show that supercell periodicity systematically biases computed octahedral tilting and energetics. Using a machine-learning interatomic potential, we compared odd-period and tilt-commensurate cells for 13 HEPOs. Odd-period cells yielded smaller and less coherent tilts and higher energies, even after configurational averaging. In ~1,700-atom cells, tilt differences persisted in all compositions and energy differences in most. The bias extended to defects, with lower oxygen vacancy formation energies in odd-period cells for strongly tilted compositions. Based on these findings, a simple design rule for equimolar supercells is proposed to lift this constraint.

## Introduction

High-entropy oxides (HEOs), in which multiple cation species occupy the same crystallographic site in nearly equal amounts, exhibit properties that differ from those of simple oxides, owing to the diverse local environments created by cation disorder and the accompanying lattice distortions and changes in electronic structure[1,2]. This concept has been extended to $ABO_3$ perovskites[3], giving rise to high-entropy perovskite oxides (HEPOs). HEPOs have attracted attention as new materials for a wide range of applications, including dielectrics[4,5], catalysts[6,7] and batteries[8,9], as summarized in a recent review[10]. Calculations based on density functional theory (DFT) and machine-learning interatomic potentials (MLIPs) play an important role in exploring their vast compositional space and in understanding their complex structures and properties[11,12]. In recent years, universal MLIPs have been increasingly applied to HEPOs, making it possible to model cells of several hundred atoms or more and many cation configurations[13–15].

These calculations often use equimolar supercells to mimic random solid solutions, constructed with methods such as the special quasirandom structure (SQS) approach[16,17]. To represent an equimolar composition, however, the number of formula units in the supercell must be a multiple of the number of components. For example, when five cation species occupy the A or B site in equal amounts, the number of formula units is restricted to multiples of five. As a result, small cells of simple shape, such as 2 × 2 × 5 (20 formula units), 3 × 3 × 5 (45 formula units) and 5 × 5 × 5 (125 formula units), tend to have an odd number of repeats in at least one direction. We refer to such cells as odd-period cells. Odd-period cells have indeed been widely used in HEPO calculations to evaluate energies related to stability and oxygen vacancies, descriptors such as lattice distortions and bond angles, and electronic structures[15,18–24].

However, odd-period cells constrain octahedral tilting, an important structural distortion in perovskites. Octahedral tilting arises from the cooperative rotation of corner-sharing $BO_6$ octahedra and influences properties such as the electronic structure, ferroelectricity and magnetism[25–27]. Because the octahedra share corners, neighboring octahedra rotate in opposite senses within the plane perpendicular to the rotation axis[28]. Placing an odd number of octahedra along a direction in this plane therefore brings octahedra rotating in the same sense next to each other across the periodic boundary, breaking the alternating arrangement (Fig. 1). The need to make the cell periodicity commensurate with octahedral tilting has been recognized for SQSs of $Pb(Zr,Ti)O_3$ solid solutions[29], simple $ABO_3$ perovskites[30,31], halide perovskites[32,33] and point-defect calculations[34].

In equimolar HEPOs with cation disorder, however, it is unclear to what extent the effect of this constraint persists. Each octahedron can distort locally in response to its surrounding cations[13,18], which may relieve the mismatch between the cell and the tilts. Alternatively, the effect may spread over long distances through the network of corner-sharing octahedra. Moreover, because different cation arrangements yield different energies[15,20], comparisons of only a few configurations cannot easily separate the effects of cell periodicity and cation configuration. It is therefore necessary to examine whether local tilting is accompanied by cell-wide coherence, how large the resulting differences in structure and energy are, and whether these differences become sufficiently small upon configurational averaging or with increasing cell size. These questions bear on the reliability of HEPO calculations, but to our knowledge they have not been examined systematically.

To address these questions, we relaxed up to 100 structures with different cation configurations per composition and cell using an MLIP and statistically evaluated how cell commensurability affects structures and energies. Specifically, for 13 HEPOs with five cation species on the A or B site, we compared odd-period cells with cells commensurate with the octahedral tilts, which accommodate all standard tilt patterns (Glazer tilt systems[28]), and examined the dependence on cell size from ~200 to ~1,700 atoms. In ~200- and ~600-atom cells, odd-period cells yielded smaller and less coherent tilts and higher energies per atom than commensurate cells in all 13 compositions. These differences decreased with increasing cell size but persisted in ~1,700-atom cells, for the tilts in all compositions and for the energy in most. We further examined how cell commensurability affects defect properties, focusing on oxygen vacancies, which are linked to the catalytic[11,24,35] and battery[8,9] performance and the oxygen nonstoichiometry[13,36] of HEPOs. In strongly tilted compositions, odd-period cells yielded systematically lower oxygen vacancy formation energies. Based on these results, we propose a supercell design rule that preserves the equimolar composition while accommodating all standard tilt patterns. This rule provides guidance for evaluating structure–property relationships across the vast compositional space of HEPOs.

## Results

### Tilt-commensurate equimolar supercells

We first summarized the conditions under which a supercell can periodically accommodate octahedral tilts. As shown in Fig. 1a, a tilt pattern can be described by the sequence of rotation senses of neighboring octahedra. Following Glazer's classification, we consider tilts with repeat periods of at most two octahedra[28]. When octahedra adjacent along the rotation axis rotate in the same sense, the tilt is called in-phase; when they rotate in opposite senses, it is called antiphase. These tilts are combinations of distortions at the R point, which corresponds to antiphase tilts, and at the three M points, which correspond to in-phase tilts about each axis[37]. To accommodate these tilts periodically, the supercell periodicity must be commensurate with the tilt patterns[33]. Hereafter, cells that can accommodate all 15 tilt systems identified by group-theoretical analysis[37] are called commensurate cells, and all other cells are called non-commensurate cells. As shown in Fig. 1b, the alternating rotations in the *ab* plane cannot continue across the periodic boundary in the $5 \times 5 \times 5$ non-commensurate cell. In the $\sqrt{20} \times \sqrt{20} \times 6$ commensurate cell, by contrast, this alternating arrangement continues periodically.

Next, we combined the commensurability condition with the constraint imposed by the equimolar composition of HEPOs. Accommodating all these tilt patterns requires every component of the three lattice vectors defining the supercell to be an even integer, in units of the lattice constant $a$ of the ideal cubic $ABO_3$ structure. The derivation is given in Eq. (1) in Methods. Commensurability requires the number of formula units to be a multiple of 8, whereas an equimolar composition with $N$ components requires it to be a multiple of $N$. The number of formula units in a cell satisfying both conditions is therefore restricted to multiples of the least common multiple of 8 and $N$. For five-component equimolar compositions, at least 40 formula units, that is, 200 atoms, are required.

We then examined the effects of cell commensurability in HEPO calculations. We studied eight compositions with five cation species in equal amounts on the A site (A-site HEPOs) and five compositions with five cation species in equal amounts on the B site (B-site HEPOs). All 13 compositions have been experimentally reported as single-phase perovskites[15,38]. As reference materials, we used $CaTiO_3$, which is experimentally known to exhibit long-range octahedral tilting[39], and $BaZrO_3$, which is known not to[40]. Hereafter, the A-site HEPOs are labeled A1–A8 and the B-site HEPOs B1–B5 in order of increasing tolerance factor $t$ (Eq. (4) in Methods). The tolerance factor is a geometric measure of compatibility derived from ionic radii[41]. When $t$ is below 1, that is, when the A cation is small relative to the octahedral framework, the octahedra tend to tilt[42], whereas when $t$ is close to or above 1, tilting becomes less likely[43]. The compositions are listed in Table S1, and the selection criteria are described in Methods.

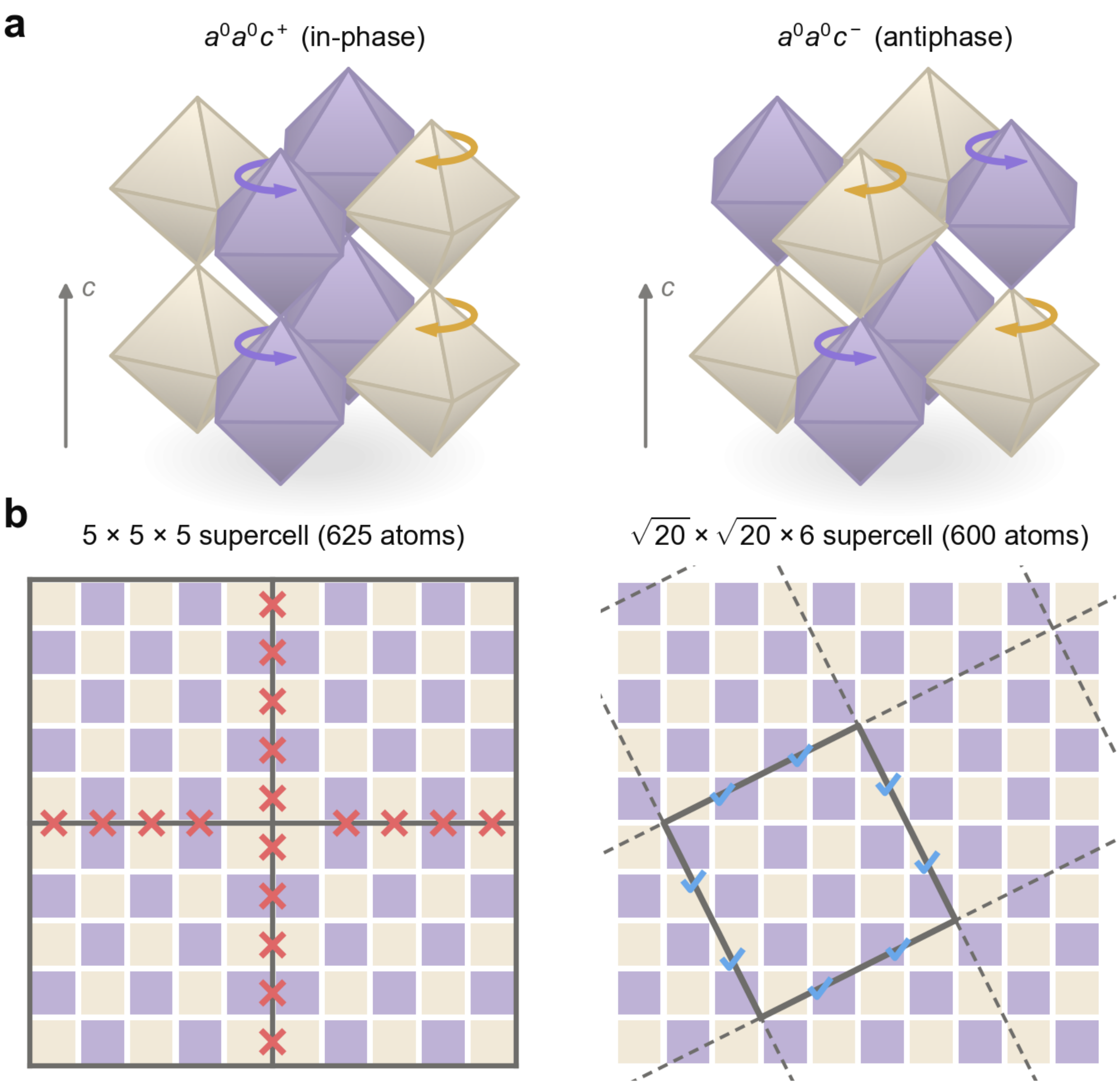

**Fig. 1 Commensurability between supercell periodicity and octahedral tilting.**

**a** Octahedral tilts about the *c* axis in Glazer notation. Corner-sharing $BO_6$ octahedra are shown schematically for two layers stacked along the *c* axis. Lavender and cream indicate counterclockwise and clockwise rotations, respectively, viewed from the positive *c* direction. Within a layer, neighboring octahedra always rotate in opposite senses. In the in-phase tilt $a^0a^0c^+$, the upper layer repeats the same pattern, whereas in the antiphase tilt $a^0a^0c^-$ the pattern is reversed. **b** Rotation senses in the *ab* plane of the non-commensurate 5 × 5 × 5 cell and the commensurate $\sqrt{20} \times \sqrt{20} \times 6$ cell. Colors are as in **a**. Gray lines mark the cell boundaries, and the cells are repeated periodically. Dashed lines in the right panel mark the boundaries of neighboring cells. In the 5 × 5 × 5 cell, the odd period places octahedra rotating in the same sense on both sides of the boundary (red ×). In the $\sqrt{20} \times \sqrt{20} \times 6$ cell, the in-plane lattice vectors (4, 2) and (−2, 4) connect octahedra rotating in the same sense, so the alternating arrangement continues across the boundary (blue ✓).

The cells compared are listed in Table 1. In the cell names, N and C denote non-commensurate and commensurate cells, respectively, and the number gives the number of atoms. The main comparison was made between the 5 × 5 × 5 N625 cell and the $\sqrt{20} \times \sqrt{20} \times 6$ C600 cell. N625 is the smallest cubic cell that can represent a five-component equimolar composition, and C600 is the commensurate cell closest to N625 in atom count. The two cells contain 25 and 24 atoms of each element on the disordered site, respectively. To examine the size dependence, we also compared the ~200-atom N225 and C200 cells and the ~1,700-atom N1575 and C1800 cells.

To handle different configurations and large cells, we performed structural relaxations with the universal MLIP Crystal Hamiltonian Graph Neural Network (CHGNet)[44]. CHGNet has been applied to rock-salt HEOs[45,46] and HEPOs[14,15] and has shown good agreement with DFT. For each composition and cell, we generated 100 SQSs for the ~200- and ~600-atom cells and 20 SQSs for the ~1,700-atom cells, and relaxed the atomic positions together with the cell shape and volume. The accuracy of CHGNet was validated against DFT calculations of ~200-atom cells, as described below.

**Table 1 Supercells compared in this study.**

Lattice vectors are given in units of the lattice constant *a* of the ideal cubic $ABO_3$ structure. N and C denote non-commensurate and commensurate cells, respectively.

| Cell | Lattice vectors | Formula units | Atoms | Accommodated R- and M-point tilts |
|---|---|---|---|---|
| N225 | (3,0,0), (0,3,0), (0,0,5) | 45 | 225 | None |
| C200 | (0,2,2), (2,2,−2), (−4,2,0) | 40 | 200 | All |
| N625 | (5,0,0), (0,5,0), (0,0,5) | 125 | 625 | None |
| C600 | (4,2,0), (−2,4,0), (0,0,6) | 120 | 600 | All |
| N1575 | (5,0,0), (0,7,0), (0,0,9) | 315 | 1,575 | None |
| C1800 | (6,0,0), (0,6,0), (0,0,10) | 360 | 1,800 | All |

## Octahedral tilt patterns in commensurate and non-commensurate cells

We examined how octahedral rotations are arranged within commensurate and non-commensurate cells. Figure 2 shows the rotation angle of each octahedron about the *b* axis on a single *ab* plane of the CHGNet-relaxed ~600-atom N625 and C600 cells. In addition to the reference materials $BaZrO_3$ and $CaTiO_3$, A2 and B2 are shown as representative HEPOs, each with one configuration per cell.

$BaZrO_3$ did not tilt in either cell and remained nearly untilted even when relaxed from a structure with octahedra tilted by 5° (Table S2). Hence, commensurability itself does not appear to induce tilting. In $CaTiO_3$, neighboring octahedra in the commensurate C600 cell rotated in opposite senses across the entire plane, so that the rotation senses formed a checkerboard pattern. In the non-commensurate N625 cell, by contrast, this checkerboard pattern did not extend over the whole plane. This result is consistent with previous reports that cell periodicity constrains the arrangement of octahedral tilts in simple perovskites[30,31]. Interestingly, this constraint imposed by cell periodicity also appeared in HEPOs with strong cation disorder. Although the rotation magnitude varied from octahedron to octahedron, the rotation senses in C600 formed a checkerboard pattern across the plane. In N625, by contrast, regions appeared in which neighboring octahedra rotated in the same sense, disrupting the checkerboard pattern. These results suggest that cell-wide coherent octahedral tilting can arise even in HEPOs composed of cations with different ionic radii and valences, but cannot be fully captured in non-commensurate cells.

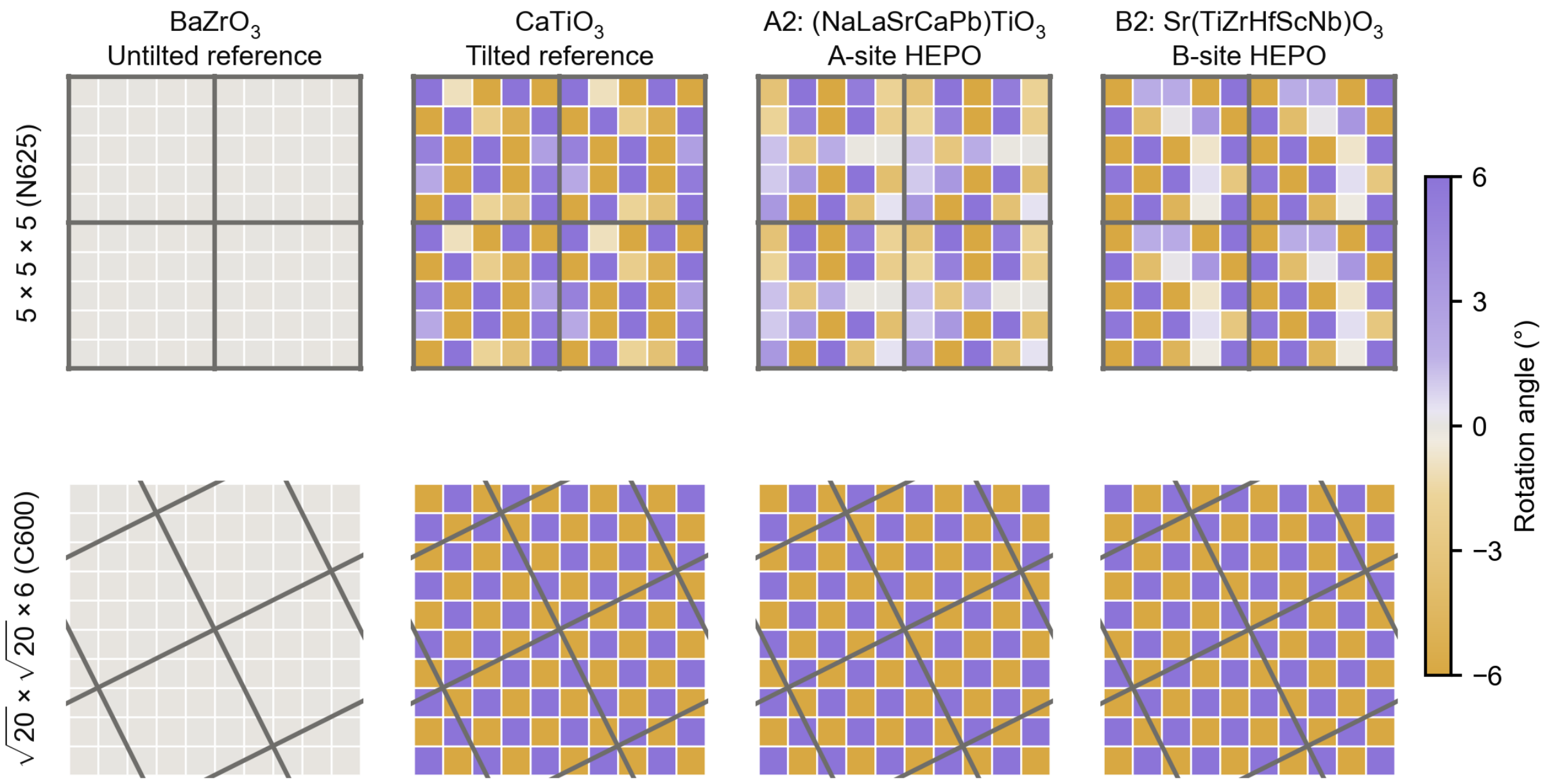

**Fig. 2 Octahedral rotations in ~600-atom commensurate and non-commensurate cells.**

Octahedra on an *ab* plane of CHGNet-relaxed structures are colored by their rotation angle about the *b* axis. The top row shows the non-commensurate N625 = 5 × 5 × 5 cell, and the bottom row the commensurate C600 = $\sqrt{20} \times \sqrt{20} \times 6$ cell. $BaZrO_3$ and $CaTiO_3$ are shown as references. A2 and B2 show the largest differences in tilt coherence $F$ between the two cells among the A-site and B-site HEPOs, respectively. For each HEPO and each cell, the configuration with $F$ closest to the median $F$ of the 100 SQSs is shown. $F$ is defined in Fig. 3 and Eq. (3) in Methods. Gray lines mark the cell boundaries, and angles beyond ±6° are shown in the end colors of the color bar.

## Configuration-averaged tilt and energy differences

Next, we examined whether the difference in tilt coherence seen in Fig. 2 persists across cation configurations. For the 100 configurations of each HEPO composition and cell, we Fourier-transformed each component of the octahedral rotation field (the rotation about each axis). We defined the tilt coherence $F$ as the sum of the intensities of the strongest waves, one from each component, divided by the total rotation intensity. $F$ equals 1 when each component is described by a single wave and decreases as the intensity spreads over multiple waves. Details are given in Eq. (3) in Methods. Figure 3a illustrates the concepts of $F$ and tilt magnitude, and Fig. 3b–d compares $F$, the tilt magnitude and the energy difference between the commensurate C600 and non-commensurate N625 cells for the 13 compositions and $CaTiO_3$. Values for each composition are listed in Table S2.

As shown in Fig. 3b, the mean $F$ over 100 configurations ranged from 0.58 to 0.84 in C600 and from 0.27 to 0.34 in N625. For all 13 compositions, the mean $F$ in N625 was no more than half that in C600. These differences far exceeded the standard deviations across configurations, indicating that the octahedral tilts were more coherent across the cell in the commensurate cells. Nevertheless, $F$ in N625 was more than three times the value of $F \approx 0.08$ obtained when the rotations were randomly shuffled among the octahedra, showing that some order in the rotation pattern remained even in the non-commensurate cells. This is consistent with the partial checkerboard patterns observed for N625 in Fig. 2. Moreover, whereas $F$ in C600 varied with composition, $F$ in N625 was nearly the same for all compositions, including $CaTiO_3$ without cation disorder. Non-commensurate cells may therefore mask compositional differences in tilt coherence.

To examine the tilt magnitude of individual octahedra, we then evaluated the root-mean-square rotation angle $\theta_{rms}$ and the B–O–B angle between adjacent octahedra. As shown in Fig. 3c, the mean $\theta_{rms}$ over 100 configurations ranged from 4.7° to 12.2° in C600 and from 3.1° to 10.5° in N625 for the 13 compositions; $\theta_{rms}$ was 1.2°–2.3° smaller in N625. The deviation of the mean B–O–B angle from 180° ranged from 12.3° to 21.2° in C600 and from 11.4° to 19.1° in N625, and was 0.9°–3.2° smaller in N625. Thus, non-commensurate cells reduced not only the cell-wide coherence of the tilts but also the tilt magnitude of individual octahedra. Among the eight A-site HEPOs, which span a range of $t$, $\theta_{rms}$ decreased with increasing $t$ in both cells (Spearman's rank correlation coefficient $\rho = -0.93$ and $-0.95$ for C600 and N625, respectively). In contrast, the correlation between $F$ and $t$ was weaker ($\rho = -0.26$ and $-0.55$, respectively). This suggests that tilt magnitude and tilt coherence are distinct properties.

As shown in Fig. 3d, the mean energy over 100 configurations was higher in N625 for all 13 compositions. The energy difference $\Delta E$ ranged from +0.16 to +5.87 meV atom$^{-1}$ for the A-site HEPOs and from +4.30 to +7.34 meV atom$^{-1}$ for the B-site HEPOs. Across the 13 compositions, the rank correlation between $\Delta E$ and $t$ was $\rho = -0.96$, consistent with reports that the energy lowering due to tilting decreases as $t$ approaches 1 and becomes small for $t \geq 1$[30,43]. Notably, $\Delta E$ decreased with increasing $t$, whereas the differences in $\theta_{rms}$ and $F$ between the cells remained. The presence of a structural constraint imposed by cell periodicity must therefore be distinguished from the size of the resulting energy difference. A small energy difference between cells alone does not mean that the two cells yield similar octahedral tilts. $CaTiO_3$ is a reference material experimentally known to exhibit long-range octahedral tilting. With one structure relaxed in each cell, the differences in $F$ and $\theta_{rms}$ between the two cells and $\Delta E$ were all larger than those for the HEPOs. Because $CaTiO_3$ has the same $t$ (0.966) as B2 and B3, the difference from the HEPOs may arise from the presence or absence of cation disorder.

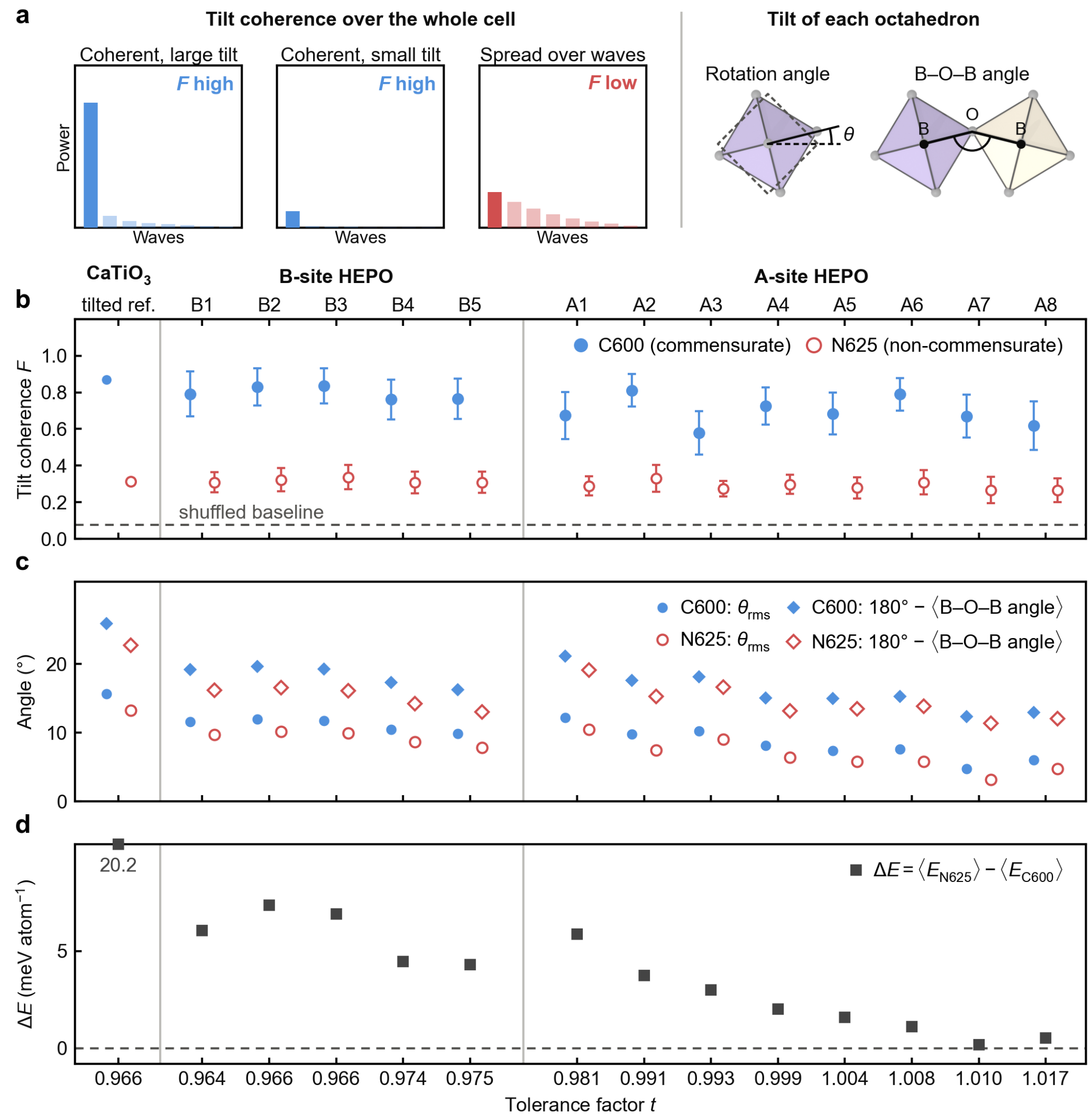

**Fig. 3 Octahedral tilting and energy in commensurate and non-commensurate cells.**

CHGNet-relaxed C600 = $\sqrt{20} \times \sqrt{20} \times 6$ and N625 = 5 × 5 × 5 cells are compared. Values for the HEPOs are for 100 configurations per composition and cell, and those for $CaTiO_3$ are for one structure per cell. In **b**–**d**, the HEPOs are placed at equal intervals, B-site HEPOs first and then A-site HEPOs, each in order of increasing tolerance factor $t$. In **c** and **d**, error bars are omitted because the standard errors are comparable to or smaller than the symbols. **a** Schematic of $F$ and tilt magnitude. The three panels on the left show the intensities of the waves obtained by decomposing the rotation field, in order of decreasing intensity. In these single-component schematics, $F$ is the fraction of the total intensity carried by the strongest wave and does not reflect the tilt magnitude itself. For the actual three-component field, $F$ is the sum of the strongest-wave intensities, one from each component, divided by the total intensity over all three components. The right panel shows the octahedral rotation angle $\theta$ and the B–O–B angle. **b** Tilt coherence $F$ in the two cells. Points and error bars for the HEPOs indicate the mean and standard deviation over 100 configurations. The dashed line shows the mean $F$ obtained when the rotation vectors are randomly shuffled among the octahedra. **c** Tilt magnitudes in the two cells. Circles show the root-mean-square rotation angle $\theta_{rms}$, and diamonds show 180° minus the mean B–O–B angle. **d** Mean energy difference per atom between the two cells, $\Delta E$ (N625 − C600). The value for $CaTiO_3$ lies outside the axis range and is given at the top.

## Supercell-size dependence of tilt and energy differences

Modeling HEPOs requires capturing the diverse local environments created by multiple cation species. Small cells can contain only a limited variety and number of local environments and may not adequately represent a random solid solution, which makes larger supercells important. We therefore examined how the structural and energy differences between commensurate and non-commensurate cells change with cell size. Figure 4a shows the three pairs of cells compared, at ~200, ~600 and ~1,700 atoms. As shown on the left of Fig. 4b,c, $\theta_{rms}$ in the commensurate cells changed by less than 0.6° from 200 to 1,800 atoms for all 13 compositions. In the non-commensurate cells, by contrast, $\theta_{rms}$ increased by 1.4°–3.2° from 225 to 1,575 atoms for all 13 compositions. This result suggests that the tilts of individual octahedra are more strongly suppressed in smaller non-commensurate cells. The $\theta_{rms}$ difference between the cells decreased with increasing cell size, from 2.5°–4.6° at ~200 atoms, but 0.7°–1.5° remained even at ~1,700 atoms. As shown in Fig. S1, $F$ was also smaller in the non-commensurate cells for all 13 compositions at all three sizes; even at ~1,700 atoms, $F$ was 0.21–0.34 in the non-commensurate cells, compared with 0.38–0.65 in the commensurate cells. Increasing the atom count alone is therefore insufficient; the cell must also be commensurate with the octahedral tilts.

As shown on the right of Fig. 4b,c, the energy difference $\Delta E$ between the cells decreased with increasing cell size for all 13 compositions. For example, $\Delta E$ decreased from 10.2 meV atom$^{-1}$ at ~200 atoms to 3.9 meV atom$^{-1}$ at ~1,700 atoms for A1, and from 13.7 to 4.3 meV atom$^{-1}$ for B2. However, $\Delta E$ persisted at ~1,700 atoms for 11 compositions, all except A7 and A8. This suggests that even large cells containing diverse local environments do not necessarily eliminate the energy difference arising from cell periodicity.

To validate the CHGNet results, we relaxed one configuration per composition and cell with DFT for the 13 compositions at ~200 atoms and compared the results with those of CHGNet for the same configurations (Fig. S2). DFT also gave smaller $F$ and $\theta_{rms}$ and higher energies in the non-commensurate cells than in the commensurate cells for all 13 compositions. The mean absolute errors (MAEs) between CHGNet and DFT were 0.014 and 0.030 for $F$ in the commensurate and non-commensurate cells, respectively, and 0.14° and 0.28° for $\theta_{rms}$, respectively. $\Delta E$ ranged from 0.51 to 16.9 meV atom$^{-1}$ with DFT and from 0.12 to 13.7 meV atom$^{-1}$ with CHGNet, with an MAE of 1.34 meV atom$^{-1}$. The rank correlations between $\Delta E$ and $t$ also agreed well ($\rho = -0.97$ for DFT and $-0.95$ for CHGNet). CHGNet gave a smaller $\Delta E$ than DFT for all 13 compositions; for the ~200-atom configurations examined here, CHGNet thus underestimated the energy difference between the cells.

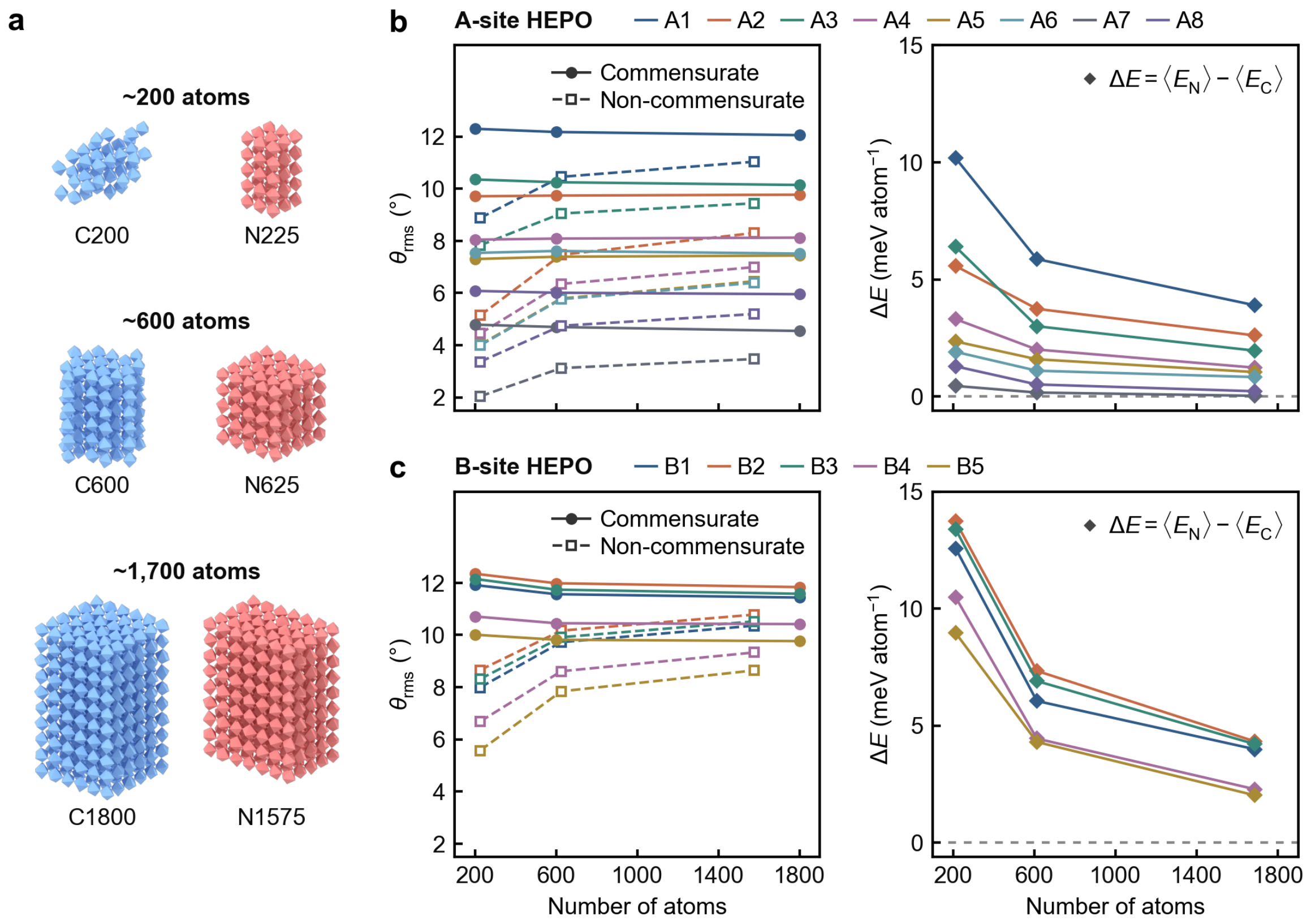

**Fig. 4 Supercell-size dependence of the octahedral tilt magnitude and the energy difference between cells.**

CHGNet-relaxed structures are compared. Values are based on 20 configurations per composition and cell for C1800 and N1575, and 100 for the other cells. **a** The six cells with different numbers of atoms compared in this study. $BO_6$ octahedra of relaxed B2 structures are drawn in the same orientation and at the same scale. Blue and red indicate commensurate and non-commensurate cells, respectively. **b**, **c** Tilt magnitude and energy difference as functions of the number of atoms for the A-site HEPOs (**b**) and the B-site HEPOs (**c**). The left panels show the configurational average of the tilt magnitude $\theta_{\mathrm{rms}}$, with solid lines and circles for commensurate cells and dashed lines and squares for non-commensurate cells. The right panels show the energy difference $\Delta E$ (non-commensurate minus commensurate) between the cells, plotted against the mean number of atoms of the two cells compared. Lines are guides to the eye.

## Effect of supercell commensurability on oxygen vacancy formation energies

We next examined whether the differences between the cells observed for defect-free structures also appear in oxygen vacancy formation energies. For each of the 13 compositions, we used 100 configurations of each ~600-atom cell, C600 and N625. We created single oxygen vacancies and relaxed the resulting structures, examining a total of 3,000 oxygen sites per composition and cell. Computational details are given in Methods. This cell size is comparable to the 625–1,280-atom cells reported in DFT and MLIP studies of oxygen vacancies in $SrTiO_3$ and HEPOs[13,47,48]. The oxygen vacancy formation energy $E_v$ is given by Eq. (5) in Methods, and its difference between cells of the same composition does not depend on the oxygen chemical potential $\mu_O$. We therefore evaluated $E_v - \mu_O$ and its difference between the cells, $E_v$(N625) − $E_v$(C600). Figure 5a shows examples of the structures of B2 in the two cells. In C600, the rotations remained alternating after oxygen removal, whereas in N625, octahedra rotating in the same sense were adjacent to each other. It has been pointed out that odd-period cells can artificially hinder the local distortions caused by defects[34], and in cubic $SrTiO_3$, the octahedral rotations induced by a vacancy are restricted by the boundary conditions of the cell[47]. In HEPOs as well, the cell periodicity may therefore affect the relaxation after oxygen removal and the formation energy.

For each composition, we compared the distributions of $E_v - \mu_O$ over the 3,000 sites in each cell. In compositions with small $t$, the distribution in the non-commensurate cell was shifted as a whole toward lower energies relative to that in the commensurate cell (Fig. S3). As representative cases, we show A7, with almost no difference, and B2, with a clear difference. For A7 ($t$ = 1.010), the distributions for the two cells nearly overlapped, with means of 8.43 eV in C600 and 8.44 eV in N625 (Fig. 5b). For B2 ($t$ = 0.966), in contrast, the entire distribution in N625 was shifted to lower energies, with a mean of 9.09 eV, 0.36 eV lower than the 9.45 eV in C600 (Fig. 5c). Across the 13 compositions, the difference in oxygen vacancy formation energy between the cells became larger in magnitude as $t$ decreased ($\rho$ = 0.98 between the difference and $t$; Fig. 5d). Among the HEPOs, the difference reached −0.38 eV and was nearly zero for compositions with $t \geq 1.008$. The reference materials without cation disorder showed the same trend, with differences of −0.41 eV for $CaTiO_3$ and −0.002 eV for $BaZrO_3$. This trend is therefore unlikely to be specific to cation disorder. For both the oxygen vacancy formation energy and the energy of the defect-free structure, the difference between the cells was larger in compositions with small $t$. However, the differences had opposite signs. In compositions with small $t$, the non-commensurate cell gave a higher energy for the defect-free structure but a lower oxygen vacancy formation energy than the commensurate cell. This suggests that the energy difference between the cells in the defect-free structures is not simply carried over to the structures after oxygen removal.

In HEPOs, cation disorder makes the local environment around oxygen differ from site to site, so that $E_v$ has a distribution rather than a single value[13,36]. We therefore examined whether the difference between the cells arises only from particular local environments. In B-site HEPOs, there are 15 possible combinations of the species of the two B cations adjacent to the removed oxygen. For B2, for example, the mean $E_v - \mu_O$ differed by about 0.9 eV among the pairs (Fig. 5e). Nevertheless, the non-commensurate cell gave lower values for all 15 pairs, with differences ranging from −0.48 to −0.24 eV. The other four B-site compositions behaved similarly. Across all five compositions, the differences between the pair-specific means ranged from −0.50 to −0.20 eV (Fig. S4). The standard errors of these differences were at most 0.05 eV, much smaller than the differences themselves. Thus, the difference between the cells in B-site HEPOs is not due to particular pairs alone.

For compositions with particularly small $t$, the difference persisted in larger cells. When the 13 compositions were compared in ~1,700-atom cells using 20 configurations and 30 sites per configuration for each composition and cell, the differences for the six compositions with $t \leq 0.981$ ranged from −0.28 to −0.16 eV, all with magnitudes greater than twice their standard errors (Fig. S5). For the seven compositions with larger $t$, the differences were small, ranging from −0.04 to +0.04 eV. Across the 13 compositions, the difference again became larger in magnitude as $t$ decreased ($\rho = 0.96$), following the same trend as at ~600 atoms. Overall, in CHGNet calculations, the choice of cell systematically changes the oxygen vacancy formation energy for compositions with small $t$, with values in non-commensurate cells up to about 0.4 eV lower than in commensurate cells.

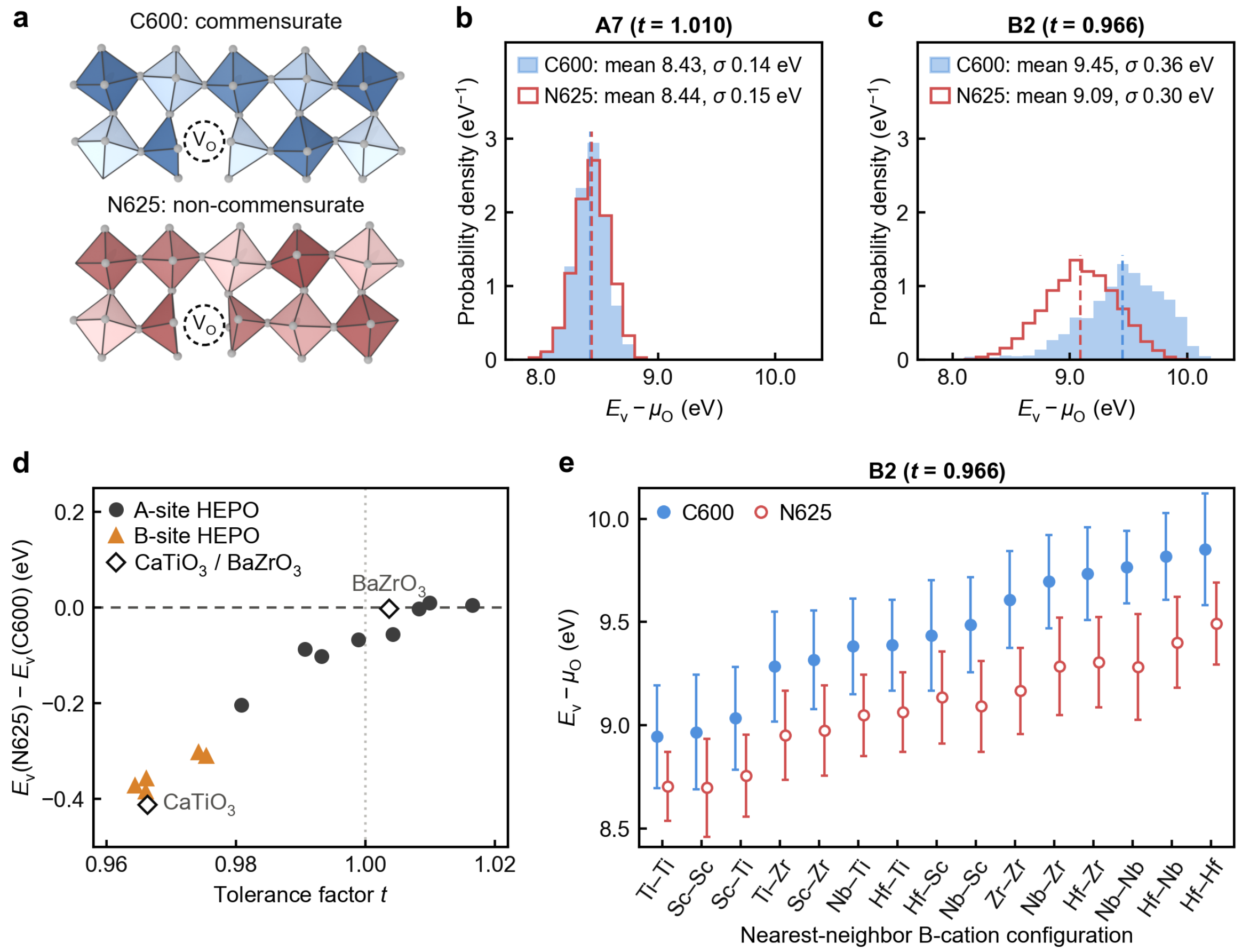

**Fig. 5 Comparison of oxygen vacancy formation energies between ~600-atom commensurate and non-commensurate cells.**

For each HEPO and cell, 30 oxygen sites were selected from each of the 100 configurations, and structures with a single oxygen vacancy at each site were relaxed with CHGNet. $E_v$ is the oxygen vacancy formation energy, and $\mu_O$ is the oxygen chemical potential. **a** Examples of relaxed B2 structures after oxygen removal. For each cell, the configuration with $F$ closest to the median $F$ of the 100 configurations is shown. Dashed circles mark the positions of the removed oxygen ($V_O$), and the color shading of the octahedra indicates the sense and magnitude of their rotation. **b**, **c** Distributions of $E_v - \mu_O$ for A7 (**b**) and B2 (**c**). Dashed lines indicate the means over all 3,000 sites, and $\sigma$ is the standard deviation. Distributions for all compositions are shown in Fig. S3. **d** Difference between the cells, $E_v$(N625) − $E_v$(C600), versus the tolerance factor $t$. For $CaTiO_3$ and $BaZrO_3$, 20 oxygen sites were examined in one structure per cell. Error bars are omitted because the standard errors, at most 0.021 eV, are comparable to the symbol size. Values for each composition are listed in Table S3. **e** $E_v - \mu_O$ of B2 for each combination of the two B-site cations nearest to the removed oxygen. Points indicate the means over all sites with that combination, and error bars indicate the standard deviations across sites. The standard errors of the differences in the means between the cells are at most 0.04 eV.

## Discussion

Non-commensurate cells are known to constrain octahedral tilting in perovskites other than HEPOs[29–34]. In equimolar HEPOs, each octahedron can adjust its tilt locally in response to cation disorder, yet the effect of cell periodicity did not disappear in this study. In odd-period cells, the coherence and magnitude of the tilts, the energy and the oxygen vacancy formation energy differed systematically from those in commensurate cells, even after averaging over many cation configurations. Even in ~1,700-atom cells, tilt differences persisted for all compositions and energy differences for most, while differences in oxygen vacancy formation energy also persisted for strongly tilted compositions. In particular, the fact that individual octahedral tilt magnitudes also decreased suggests that the constraint imposed by the periodic boundary extends to local tilting. These quantities also serve as computational indicators. Energies are used to assess compositional stability[15,18], and tilts and B–O–B angles to characterize local structure[13,36]. Oxygen vacancy formation energies are used to evaluate the catalytic activity[11,24] and to predict the oxygen nonstoichiometry[13,36] of HEPOs. Evaluations based on calculations with odd-period cells may therefore depend on the choice of cell.

The energy differences arising from the choice of cell are not negligible for assessing phase stability. For B2, which shows the largest $\Delta E$, the difference was 13.7 meV atom$^{-1}$ at ~200 atoms (16.9 meV atom$^{-1}$ with DFT), about half the magnitude of the ideal configurational entropy contribution to the free energy[49] at 1,000 K for five elements occupying the A or B site in equal amounts, $(k_B T \ln 5)/5 \approx$ 27.7 meV atom$^{-1}$. Even at ~1,700 atoms, $\Delta E$ was about 16% of this value. Although this comparison does not directly indicate a change in phase stability, the choice of cell requires care when energies obtained with DFT or MLIPs are used as indicators of phase stability.

The difference in oxygen vacancy formation energy between the cells cannot be explained solely by the difference in tilt coherence. For compositions with $t \geq 1.008$, the formation energies were nearly the same in the two cells, although $F$ differed greatly between them. Moreover, in B-site HEPOs, formation energies were lower in the non-commensurate cell for every combination of the two B cations adjacent to the removed oxygen. Thus, the difference cannot be explained solely by the identities of the nearest-neighbor B-cation pairs either.

Most of this difference in oxygen vacancy formation energy arose from relaxation after oxygen removal. In HEPOs with small $t$, the difference between the cells, which was small immediately after oxygen removal, increased in magnitude after atomic relaxation (from −0.07 to −0.36 eV for B2; Table S3). In other words, the energy lowering upon relaxation was larger in the non-commensurate cell. It has been pointed out that odd-period cells can suppress octahedral tilting[34], and this larger energy lowering may be related to the periodic constraint in the defect-free structure. Local breaking of the B–O–B connectivity upon oxygen removal may relieve part of this frustration. Effects of tilting on oxygen vacancy formation energies have also been reported for $SrTiO_3$[48] and $PbTiO_3/SrTiO_3$ superlattices[50]. However, these studies did not examine the mismatch imposed by periodic boundaries. In addition, the cells compared in this study also differ in shape and cation configuration. Clarifying the mechanism will require a more detailed analysis of the changes in local structure after oxygen removal, such as tilts and cation displacements.

The choice of cell also affects comparisons between compositions. Because the extent to which non-commensurate cells lower the oxygen vacancy formation energy differs among compositions, using the same cell for all compositions does not preserve the differences in formation energy between compositions obtained with commensurate cells. Indeed, the range (maximum minus minimum) of the mean $E_v - \mu_O$ over the 13 compositions at ~600 atoms narrowed from 1.90 eV in the commensurate cells to 1.63 eV in the non-commensurate cells. Similarly, the range of the mean $F$ over the 13 compositions narrowed from 0.26 in the commensurate cells to 0.07 in the non-commensurate cells. The choice of cell therefore affects how the compositional dependence is captured, not only for formation energies but also for tilt coherence.

For equimolar HEPOs, choosing a cell commensurate with the tilts is thus important, beyond simply increasing the number of atoms. For five-component equimolar HEPOs, accommodating all standard tilt patterns while maintaining the equimolar composition requires cells in which every lattice-vector component, expressed in units of the ideal cubic lattice constant $a$, is even and the number of formula units is a multiple of 40. Among diagonal supercells, the 200-atom $2 \times 2 \times 10$ cell is the smallest example. By contrast, cells with odd periods in only some directions, such as $2 \times 2 \times 5$, accommodate only some of the tilts, such as the in-phase tilt about the $c$ axis. Moreover, because tilt differences remain even for compositions with small energy differences, the validity of a cell should be checked not only by the energy but also by the tilt magnitude and coherence. This cell-design rule for HEPOs is expected to improve the reliability of computational screening based on energies and oxygen vacancy formation energies. Furthermore, because octahedral tilting is related to the electronic structure, ferroelectricity and magnetism[25–27], the choice of cell may also affect calculations of these properties. The design rule presented here provides a computational basis for evaluating the structural, energetic and defect properties of HEPOs.

## Methods

### Commensurability condition

We write each lattice vector of a supercell as $T = (x, y, z)$ in units of the lattice constant $a$ of the ideal cubic $ABO_3$ structure, where $x$, $y$ and $z$ are integers. With a wave vector $q$ expressed in fractional coordinates with respect to the reciprocal-lattice basis of this cubic lattice, a distortion with wave vector $q$ can be accommodated under periodic boundary conditions only if $q \cdot T$ is an integer for all lattice vectors[51]. In Glazer notation, antiphase tilts (−) correspond to the R point, $q = (½, ½, ½)$, and in-phase tilts (+) about the $a$, $b$ and $c$ axes correspond to the M points, $q = (0, ½, ½)$, $(½, 0, ½)$ and $(½, ½, 0)$, respectively[37]. All tilts at the R point and the three M points can therefore be accommodated if and only if the following four quantities are even for all lattice vectors:

$$x + y + z,\ y + z,\ z + x,\ x + y \tag{1}$$

If $y + z$, $z + x$ and $x + y$ are all even, then $x$, $y$ and $z$ have the same parity, and if all three are odd, $x + y + z$ is odd. The condition is therefore equivalent to every component of the three lattice vectors being even. In N225, N625 and N1575, the periods along all three axes are odd, so none of the R- and M-point tilts can be accommodated periodically.

### Analysis of octahedral tilting

The rotation of each $BO_6$ octahedron was determined with the Kabsch algorithm[52], which finds the rotation that best aligns the six ideal octahedral directions with the corresponding B–O directions. Each rotation was expressed as a vector $\omega_j$ whose direction gives the rotation axis and whose length gives the rotation angle. The tilt magnitude $\theta_{\mathrm{rms}}$ is the root mean square of $|\omega_j|$ over all octahedra in the cell. The B–O–B angle is the average of the angles subtended at each oxygen by its two nearest B cations.

The tilt coherence $F$ measures how strongly each component of the octahedral rotation field is concentrated in a single wave. For each component $\omega_\alpha$ of $\omega_j$ along the $a$, $b$ and $c$ axes, we computed the Fourier transform over the wave vectors $q$ allowed by the cell periodicity[53]:

$$\omega_\alpha(q) = \frac{1}{N}\sum_j \omega_\alpha(j)\exp(-2\pi i q \cdot r_j) \tag{2}$$

where $r_j$ is the lattice position of B site $j$ and $N$ is the number of octahedra. $F$ was then obtained as

$$F = \frac{\sum_\alpha \max_q W_\alpha(q)}{\theta_{\mathrm{rms}}^2} \tag{3}$$

Because the rotation field is real, $q$ and $-q$ represent the same wave, and $W_\alpha(q)$ is therefore defined as the sum of their intensities, $|\omega_\alpha(q)|^2 + |\omega_\alpha(-q)|^2$. For wave vectors at which $q$ and $-q$ are equivalent modulo a reciprocal-lattice vector of the ideal cubic lattice, such as the R and M points, $W_\alpha(q) = |\omega_\alpha(q)|^2$. $F$ equals 1 when each component is described by a single wave and decreases as the intensity spreads over multiple waves. Because the definition is not restricted to the R and M points, commensurate and non-commensurate cells can be compared using the same definition. Variations in the rotation magnitude also spread the intensity, so $F$ can be smaller than 1 even when the rotation senses form a regular pattern.

## Compositions

The 13 compositions studied are experimentally reported HEPOs, each containing five cation species in equal amounts on the A or B site. The eight A-site HEPOs were selected from compositions synthesized in a previous study[15] and confirmed by X-ray diffraction to form a single phase. The five B-site HEPOs are the compositions without oxygen deficiency among those that Shishkin and Kiryakov reported as single-phase perovskites[38]. These compositions are charge-balanced with nominal valences and therefore require no charge-compensating oxygen vacancies. Because all cations are nominally closed-shell, spin polarization was not considered in the DFT calculations. The tolerance factor $t$ is a geometric measure of compatibility calculated from ionic radii as[41]

$$t = \frac{r_{\mathrm{A}}+r_{\mathrm{O}}}{\sqrt{2}(r_{\mathrm{B}}+r_{\mathrm{O}})} \tag{4}$$

Here, $r_{\mathrm{A}}$, $r_{\mathrm{B}}$ and $r_{\mathrm{O}}$ are the ionic radii of the A-site cations, B-site cations and oxygen, respectively. For the ionic radii, we used values from the machine-learning-extended Shannon database[54,55] for the cations and the Shannon value[56] for $O^{2-}$, with coordination numbers of 12 for the A site and 6 for the B site and oxygen. For the site occupied by five cation species, the radius was taken as the mole-fraction-weighted average of the individual cation radii. As reference materials, we used $CaTiO_3$, which is experimentally known to exhibit long-range octahedral tilting, and $BaZrO_3$, which is known not to. The 13 compositions and their $t$ values are listed in Table S1, and the ionic radii used are given in Table S4.

## SQS generation and CHGNet relaxation

To represent disordered cation arrangements, SQSs[16] were generated for each composition and cell with the Integrated Cluster Expansion Toolkit (icet)[57], starting from the ideal cubic perovskite structure. The number of configurations was 100 per composition and cell for the ~200- and ~600-atom cells and 20 for the ~1,700-atom cells. The SQSs were relaxed with the pretrained universal MLIP CHGNet[44] (version 0.4.0, weights 0.3.0) using the Atomic Simulation Environment (ASE)[58]. The cell shape, volume and atomic positions were optimized simultaneously until the maximum force on any atom fell below 0.02 eV Å$^{-1}$. To break the symmetry of the ideal $CaTiO_3$ and $BaZrO_3$ structures, small random displacements (standard deviation 0.05 Å) were added to all atoms before relaxation.

## DFT validation

DFT calculations were performed with the Vienna Ab initio Simulation Package (VASP) 5.4.4[59] using the projector augmented-wave (PAW) method[60,61] and the Perdew–Burke–Ernzerhof (PBE) functional[62]. For consistency with the Materials Project calculations[63] used to train CHGNet, the input settings were based on MPRelaxSet. The plane-wave cutoff energy was 520 eV, and the electronic convergence criterion was $10^{-6}$ eV. Following the validation procedure of Sivak et al.[45], we used the CHGNet-relaxed structures as starting structures and relaxed the cell shape, volume and atomic positions until the maximum force fell below 0.02 eV $Å^{-1}$. Γ-centered k-point meshes of 3 × 3 × 2 and 3 × 2 × 2 were used for N225 and C200, respectively. For each composition and cell, one of the 100 CHGNet-relaxed configurations with an energy near the median was selected. These configurations were chosen before the DFT calculations were performed.

## Oxygen vacancy formation energies

Oxygen vacancy formation energies were calculated for the ~600-atom C600 and N625 cells using the 100 configurations per cell for each of the 13 compositions, with 30 oxygen sites randomly selected from each configuration. For $CaTiO_3$ and $BaZrO_3$, 20 sites were examined in one structure per cell. For the ~1,700-atom C1800 and N1575 cells, 20 configurations per cell and 30 sites per configuration were examined for each of the 13 compositions. The CHGNet-relaxed host supercells were first re-relaxed with the lattice fixed. One oxygen atom was then removed, and only the atomic positions were relaxed, with the lattice kept fixed[34], until the maximum force fell below 0.02 eV $Å^{-1}$. The formation energy of a neutral oxygen vacancy, $E_v$, is given by

$$E_{\mathrm{v}} = E_{\mathrm{defect}} - E_{\mathrm{bulk}} + \mu_{\mathrm{O}} \tag{5}$$

where $E_{\mathrm{defect}}$ and $E_{\mathrm{bulk}}$ are the total energies of the defect and host supercells, respectively, and $\mu_{\mathrm{O}}$ is the oxygen chemical potential. In this study, we evaluated $E_{\mathrm{v}} - \mu_{\mathrm{O}}$. Assuming that $\mu_{\mathrm{O}}$ is common to all compositions, the differences between compositions do not depend on the value of $\mu_{\mathrm{O}}$.

## Use of large language models

Claude (Anthropic, Claude Opus 5.5) was used to assist with data analysis and code development. All LLM-generated outputs were critically reviewed, verified and revised by the authors, who assume full responsibility for the final content.

## Code availability

The scripts used for the octahedral-rotation analysis are available from the corresponding author upon reasonable request.

## Data availability

The data that support the findings of this study are available from the corresponding author upon reasonable request.

## Author contributions

H.T. conceived the project, designed the study, performed the computational analyses, analyzed the data, prepared the figures, and wrote the original draft. W.K.T., H.M., A.M., T.I., K.H., and R.M. reviewed the manuscript and provided feedback. K.H. and R.M. also provided computational resources. L.K. provided computational chemistry advice and critically revised the manuscript. G.K. supervised the project, acquired funding, contributed to discussions, and critically revised the manuscript. All authors reviewed and approved the final manuscript.

## Acknowledgements

This work was supported by JSPS KAKENHI Grant Numbers JP23K26377, JP25K22154 and JP25H01668 (Grant-in-Aid for Transformative Research Areas (A) "Supra-ceramics"), the Asahi Glass Foundation, the Nitto Foundation, and a research grant from the Rinnai Scholarship Foundation. The computations in this research were partially performed using the facilities of Center for Advanced Scientific Computing at JAIST. The authors also thank the Information and Media Center (IMC) at Toyohashi University of Technology for providing computational resources through its HPC cluster.

## Competing interests

The authors declare no competing interests.

Supplementary Information

# Supercell periodicity biases octahedral tilting and energetics in high-entropy perovskite oxides

Hanzo Tsubota[1], Wai Kian Tan[2,3], Hiroyuki Muto[1,2,3], Atsunori Matsuda[1,2], Tom Ichibha[4], Kenta Hongo[5], Ryo Maezono[6], Liangzhi Kou[7], Go Kawamura[1,2*]

[1] Department of Electrical and Electronic Information Engineering, Toyohashi University of Technology, Toyohashi, Aichi 441-8580, Japan

[2] Institute for Research on Next Generation Semiconductor and Sensing Science, Toyohashi University of Technology, Toyohashi, Aichi 441-8580, Japan

[3] Institute of Liberal Arts and Sciences, Toyohashi University of Technology, Toyohashi, Aichi 441-8580, Japan

[4] School of Information Science, Japan Advanced Institute of Science and Technology, Nomi, Ishikawa 923-1292, Japan

[5] Center for Advanced Scientific Computing, Japan Advanced Institute of Science and Technology, Nomi, Ishikawa 923-1292, Japan

[6] Graduate Major in Materials and Information Sciences, School of Materials and Chemical Technology, Institute of Science Tokyo, Meguro-ku, Tokyo 152-8550, Japan

[7] School of Mechanical, Medical and Process Engineering, Queensland University of Technology, Brisbane, QLD 4001, Australia

*Corresponding author: Go Kawamura (kawamura.go.km@tut.jp)

**Table S1 The 13 compositions and their tolerance factors *t*.**

Each composition contains five cation species in equal amounts on the A or B site and has been experimentally reported as a single-phase perovskite. For the tolerance factor *t*, the radius of the mixed site was taken as the average of the individual cation radii. The mean nominal valence of the cations on the mixed site is +2 for the A-site HEPOs and +4 for the B-site HEPOs. Within each group, the compositions are numbered in order of increasing *t*.

| Name | Composition | *t* | Ref. |
|---|---|---|---|
| A1 | $(Na_{0.2}Sr_{0.2}Ca_{0.2}Pb_{0.2}Y_{0.2})TiO_3$ | 0.981 | 1 |
| A2 | $(Na_{0.2}La_{0.2}Sr_{0.2}Ca_{0.2}Pb_{0.2})TiO_3$ | 0.991 | 1 |
| A3 | $(Na_{0.2}K_{0.2}La_{0.2}Sr_{0.2}Y_{0.2})TiO_3$ | 0.993 | 1 |
| A4 | $(Na_{0.2}La_{0.2}Sr_{0.2}Ca_{0.2}Ba_{0.2})TiO_3$ | 0.999 | 1 |
| A5 | $(Bi_{0.2}Na_{0.2}K_{0.2}La_{0.2}Sr_{0.2})TiO_3$ | 1.004 | 1 |
| A6 | $(K_{0.2}La_{0.2}Sr_{0.2}Ca_{0.2}Pb_{0.2})TiO_3$ | 1.008 | 1 |
| A7 | $(Na_{0.2}La_{0.2}Sr_{0.2}Ba_{0.2}Pb_{0.2})TiO_3$ | 1.010 | 1 |
| A8 | $(K_{0.2}La_{0.2}Sr_{0.2}Ca_{0.2}Ba_{0.2})TiO_3$ | 1.017 | 1 |
| B1 | $Sr(Mg_{0.2}Zr_{0.2}Hf_{0.2}Ta_{0.2}Nb_{0.2})O_3$ | 0.964 | 2 |
| B2 | $Sr(Ti_{0.2}Zr_{0.2}Hf_{0.2}Sc_{0.2}Nb_{0.2})O_3$ | 0.966 | 2 |
| B3 | $Sr(Ti_{0.2}Zr_{0.2}Hf_{0.2}Sc_{0.2}Ta_{0.2})O_3$ | 0.966 | 2 |
| B4 | $Sr(Ti_{0.2}Zr_{0.2}Mg_{0.2}Ta_{0.2}Nb_{0.2})O_3$ | 0.974 | 2 |
| B5 | $Sr(Ti_{0.2}Mg_{0.2}Hf_{0.2}Ta_{0.2}Nb_{0.2})O_3$ | 0.975 | 2 |
| Reference material | $CaTiO_3$ | 0.966 | — |
| Reference material | $BaZrO_3$ | 1.004 | — |

**Table S2 Octahedral tilting and energy differences in the ~600-atom cells.**

For the HEPOs, values are means over 100 configurations per composition and cell for CHGNet-relaxed structures. $\theta_{rms}$ is the root-mean-square rotation angle of the $BO_6$ octahedra, $F$ is the tilt coherence, 180° − B–O–B is the deviation of the mean B–O–B angle from 180° (a linear B–O–B linkage), and $\Delta E$ is the energy difference per atom between the cells (N625 − C600). The standard deviations (SDs) of the energies over the 100 configurations of each cell are also given. Values for $CaTiO_3$ and $BaZrO_3$ are for one structure per cell. $F$ is not given for $BaZrO_3$ because it is nearly untilted. For $BaZrO_3$, $\theta_{rms}$ remained at 0.53° in C600 and 0.05° in N625 even when relaxed from a structure with octahedra tilted by 5°.

| | **$\theta_{rms}$ (°)** | | ***F*** | | **180° − B–O–B (°)** | | **Energy (meV atom$^{-1}$)** | | |
|---|---|---|---|---|---|---|---|---|---|
| | **C600** | **N625** | **C600** | **N625** | **C600** | **N625** | **$\Delta E$** | **SD C600** | **SD N625** |
| **A1** | 12.17 | 10.46 | 0.67 | 0.29 | 21.16 | 19.10 | +5.87 | 1.16 | 1.37 |
| **A2** | 9.74 | 7.47 | 0.81 | 0.33 | 17.63 | 15.24 | +3.74 | 0.45 | 0.58 |
| **A3** | 10.25 | 9.05 | 0.58 | 0.27 | 18.15 | 16.60 | +3.00 | 0.83 | 0.62 |
| **A4** | 8.09 | 6.35 | 0.73 | 0.30 | 15.05 | 13.14 | +2.00 | 0.47 | 0.48 |
| **A5** | 7.39 | 5.79 | 0.68 | 0.28 | 15.00 | 13.45 | +1.59 | 0.63 | 0.59 |
| **A6** | 7.61 | 5.77 | 0.79 | 0.31 | 15.27 | 13.87 | +1.10 | 0.48 | 0.71 |
| **A7** | 4.70 | 3.12 | 0.67 | 0.27 | 12.30 | 11.39 | +0.16 | 0.50 | 0.61 |
| **A8** | 6.01 | 4.75 | 0.62 | 0.27 | 12.94 | 12.02 | +0.51 | 0.46 | 0.63 |
| **B1** | 11.56 | 9.72 | 0.79 | 0.31 | 19.18 | 16.17 | +6.05 | 1.93 | 2.04 |
| **B2** | 11.98 | 10.17 | 0.83 | 0.32 | 19.66 | 16.58 | +7.34 | 0.81 | 0.50 |
| **B3** | 11.74 | 9.91 | 0.84 | 0.34 | 19.22 | 16.12 | +6.90 | 0.61 | 0.57 |
| **B4** | 10.45 | 8.61 | 0.76 | 0.31 | 17.28 | 14.22 | +4.45 | 1.99 | 1.74 |
| **B5** | 9.82 | 7.84 | 0.76 | 0.31 | 16.28 | 13.05 | +4.30 | 2.09 | 2.19 |
| **$CaTiO_3$** | 15.65 | 13.25 | 0.87 | 0.31 | 25.87 | 22.67 | +20.24 | — | — |
| **$BaZrO_3$** | 0.01 | 0.00 | — | — | 0.02 | 0.01 | +0.00 | — | — |

**Table S3 Oxygen vacancy formation energies in the ~600-atom cells.**

Values are $E_v - \mu_O$ (eV) calculated with CHGNet. For the HEPOs, 30 sites were examined in each of the 100 configurations per cell, and the means are taken over the 3,000 sites of each cell. The standard errors (SEs) of the differences between the cells were calculated using the 100 configuration-level means in each cell, each averaged over 30 sites. The unrelaxed differences were obtained without relaxing the atoms after oxygen removal. For $CaTiO_3$ and $BaZrO_3$, 20 sites were examined in one structure per cell, and no SE is given.

| | $E_v - \mu_O$ (eV) | | Difference (eV) | |
|---|---|---|---|---|
| | **C600** | **N625** | **N625 − C600 ± SE** | **Unrelaxed N625 − C600 ± SE** |
| **A1** | 8.279 | 8.075 | −0.203 ± 0.020 | −0.030 ± 0.007 |
| **A2** | 8.457 | 8.370 | −0.087 ± 0.011 | +0.014 ± 0.006 |
| **A3** | 8.094 | 7.993 | −0.101 ± 0.021 | −0.018 ± 0.004 |
| **A4** | 8.413 | 8.346 | −0.067 ± 0.010 | +0.001 ± 0.004 |
| **A5** | 8.237 | 8.181 | −0.056 ± 0.011 | +0.001 ± 0.005 |
| **A6** | 8.440 | 8.437 | −0.003 ± 0.010 | +0.038 ± 0.006 |
| **A7** | 8.426 | 8.437 | +0.010 ± 0.009 | +0.018 ± 0.006 |
| **A8** | 8.398 | 8.403 | +0.005 ± 0.009 | +0.014 ± 0.005 |
| **B1** | 9.998 | 9.627 | −0.372 ± 0.017 | −0.064 ± 0.007 |
| **B2** | 9.446 | 9.088 | −0.358 ± 0.021 | −0.072 ± 0.005 |
| **B3** | 9.862 | 9.478 | −0.384 ± 0.017 | −0.083 ± 0.006 |
| **B4** | 9.564 | 9.261 | −0.303 ± 0.014 | −0.051 ± 0.006 |
| **B5** | 9.720 | 9.410 | −0.310 ± 0.013 | −0.049 ± 0.008 |
| **$CaTiO_3$** | 8.743 | 8.330 | −0.412 | −0.026 |
| **$BaZrO_3$** | 8.705 | 8.703 | −0.002 | 0.000 |

**Table S4 Ionic radii used for the tolerance factor.**

Cation radii are taken from the machine-learning-extended Shannon database[3,4], and the $O^{2-}$ radius is the Shannon value[5]. The coordination numbers are 12 for the A-site cations and 6 for the B-site cations and $O^{2-}$.

| Ion | Site | Radius (Å) | Ion | Site | Radius (Å) |
|---|---|---|---|---|---|
| $Na^{+}$ | A | 1.393 | $Bi^{3+}$ | A | 1.380 |
| $K^{+}$ | A | 1.642 | $Ti^{4+}$ | B | 0.6028 |
| $Ca^{2+}$ | A | 1.337 | $Zr^{4+}$ | B | 0.7207 |
| $Sr^{2+}$ | A | 1.445 | $Hf^{4+}$ | B | 0.7083 |
| $Ba^{2+}$ | A | 1.610 | $Sc^{3+}$ | B | 0.7411 |
| $Pb^{2+}$ | A | 1.492 | $Nb^{5+}$ | B | 0.6377 |
| $La^{3+}$ | A | 1.362 | $Ta^{5+}$ | B | 0.6390 |
| $Y^{3+}$ | A | 1.223 | $Mg^{2+}$ | B | 0.7245 |
| $O^{2-}$ | O | 1.400 | | | |

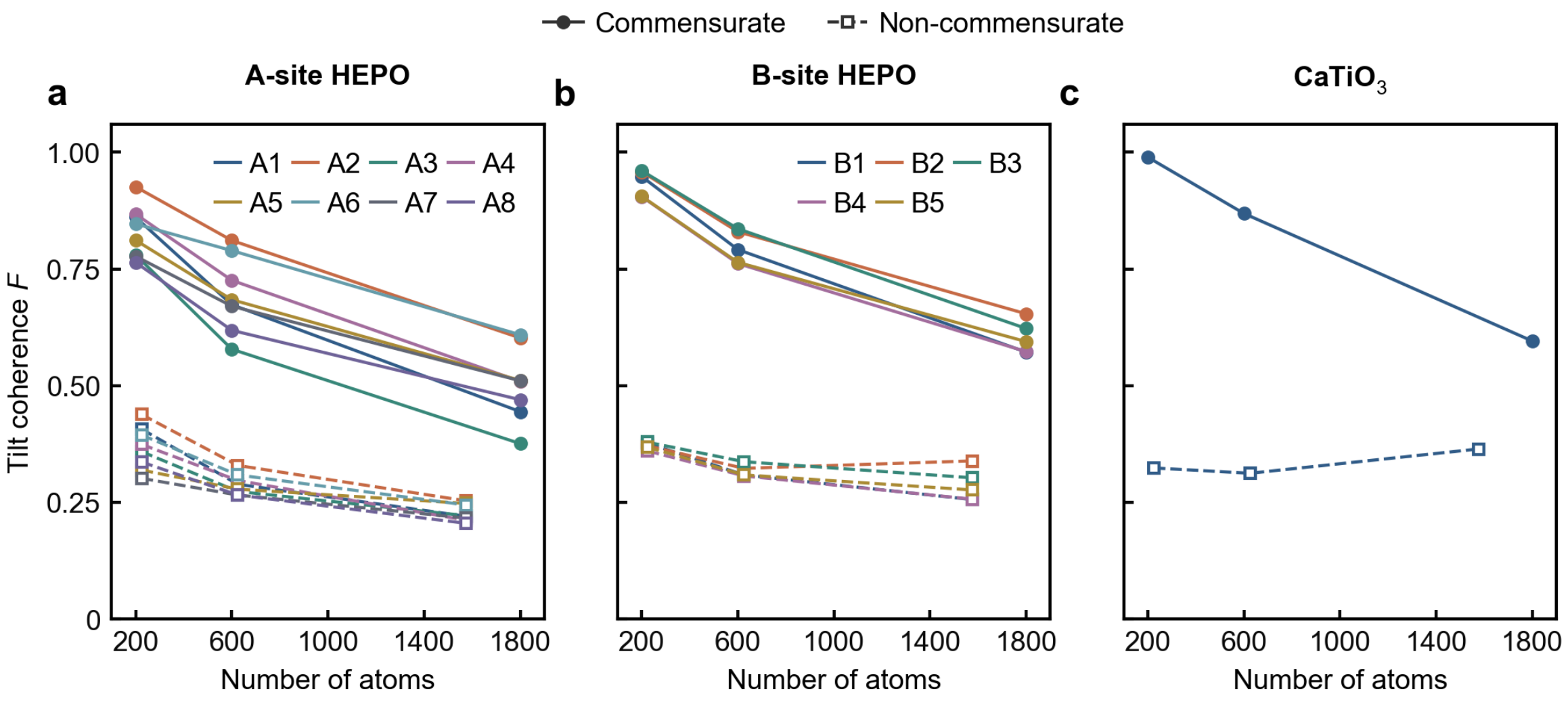

**Fig. S1 Supercell-size dependence of the tilt coherence *F* in HEPOs and $CaTiO_3$.**

CHGNet-relaxed commensurate cells (C200, C600 and C1800) and non-commensurate cells (N225, N625 and N1575) are compared. Circles with solid lines denote commensurate cells, and squares with dashed lines denote non-commensurate cells. **a** The eight A-site HEPOs. **b** The five B-site HEPOs. **c** $CaTiO_3$. Points for the HEPOs are means over 100 configurations per composition and cell for the ~200- and ~600-atom cells and over 20 configurations per composition and cell for the ~1,700-atom cells. Values for $CaTiO_3$ are for one structure per cell.

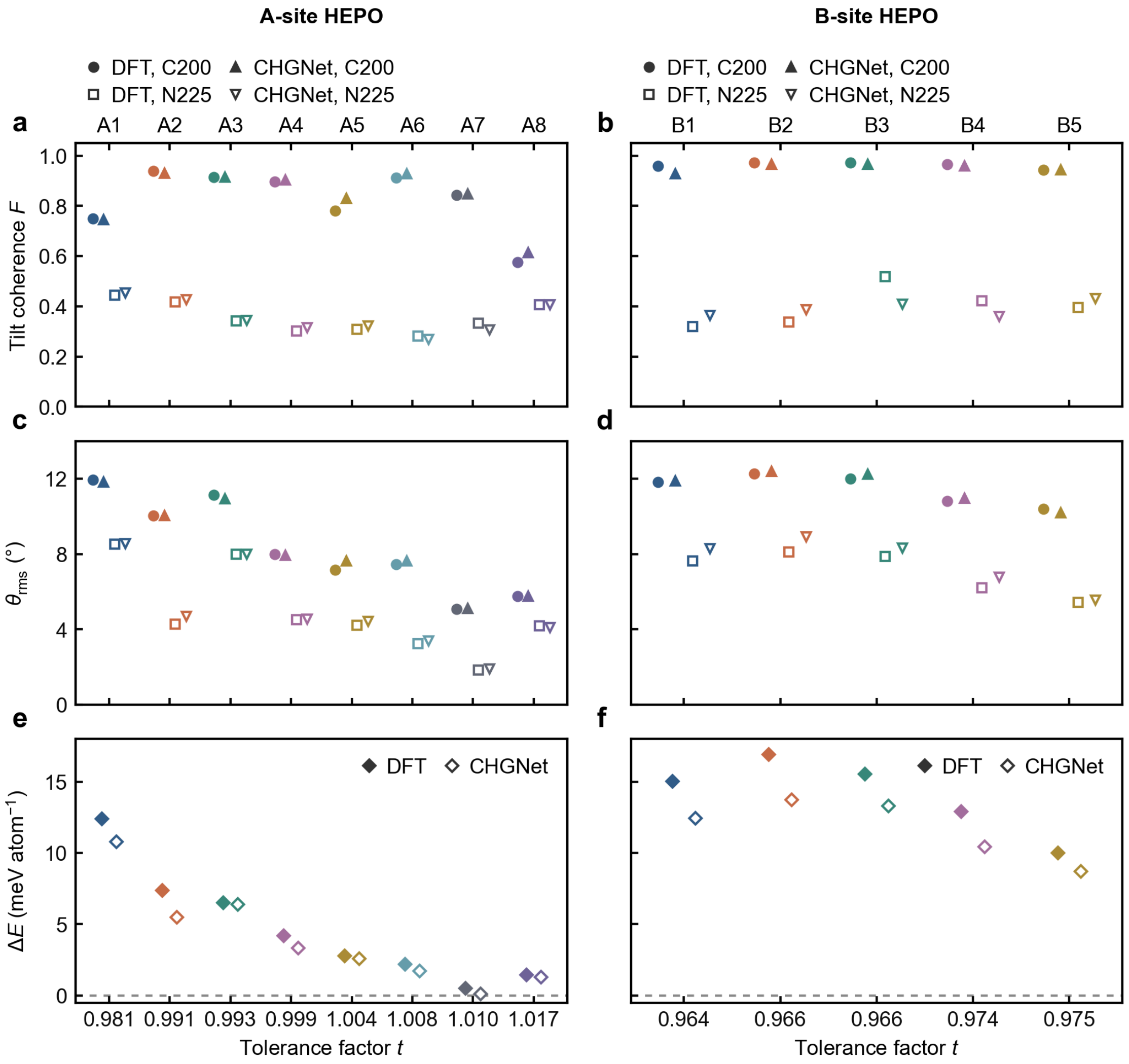

**Fig. S2 DFT and CHGNet comparison of C200 and N225 for the same cation configurations.**

The left and right columns show the A-site and B-site HEPOs, respectively. One configuration per composition and cell was used. Within each group, the compositions are placed at equal intervals in order of increasing tolerance factor $t$. **a**, **b** Tilt coherence $F$. **c**, **d** Root-mean-square rotation angle $\theta_{\mathrm{rms}}$ of the octahedra. **e**, **f** Energy difference per atom, $\Delta E = E(\mathrm{N225}) - E(\mathrm{C200})$. The same pair of cation configurations, one for C200 and one for N225, was used for both methods to calculate the energy difference. Error bars are not shown because one configuration per cell was compared.

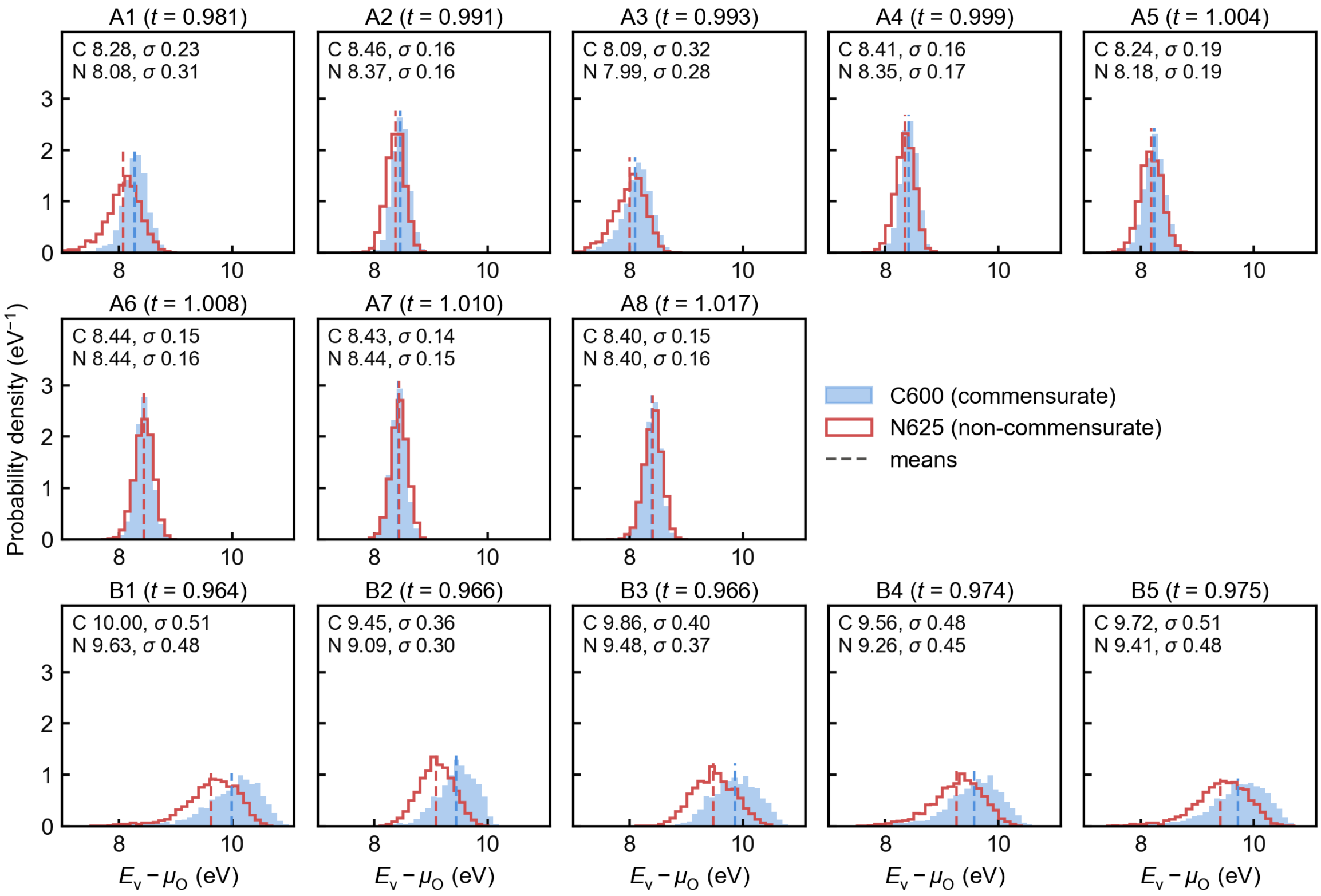

**Fig. S3 Distributions of the oxygen vacancy formation energy $E_v - \mu_O$ for the 13 compositions.**

Values for all 3,000 sites in each CHGNet-relaxed ~600-atom cell (30 sites in each of the 100 configurations) are shown. The vertical axis is the probability density. Filled histograms show the commensurate C600 cell, outlined histograms the non-commensurate N625 cell, and dashed lines the mean of each cell. The labels C and N in each panel give the mean and standard deviation $\sigma$ (eV) of each cell. The bin width is 0.1 eV, and all panels share the same axis ranges.

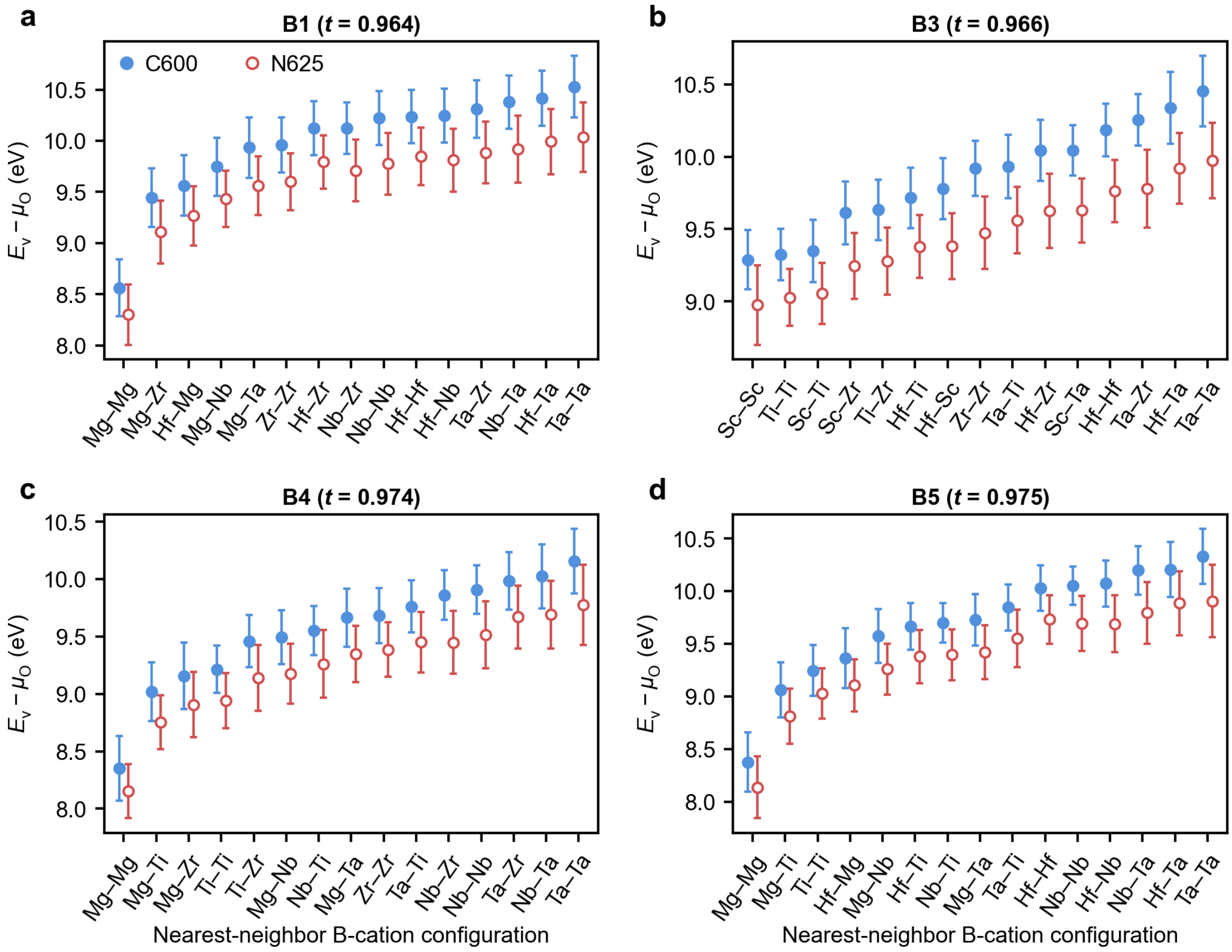

**Fig. S4 Oxygen vacancy formation energies in B-site HEPOs for each combination of the two B cations adjacent to the removed oxygen.**

$E_v - \mu_O$ for 30 sites in each of the 100 configurations per cell in the CHGNet-relaxed ~600-atom C600 and N625 cells is shown separately for each of the 15 combinations of the species of the two B-site cations nearest to the removed oxygen. **a** B1. **b** B3. **c** B4. **d** B5. B2 is shown in Fig. 5e. Points indicate the means over all sites with each combination, and error bars indicate the standard deviations across sites. The combinations are ordered by increasing C600 mean. For all 15 combinations, the mean in N625 is lower than that in C600, with differences ranging from −0.50 to −0.26 eV for B1, from −0.48 to −0.30 eV for B3, from −0.41 to −0.20 eV for B4 and from −0.42 to −0.21 eV for B5. The standard errors of the differences in the means between the cells, calculated from the per-SQS means, are at most 0.05 eV.

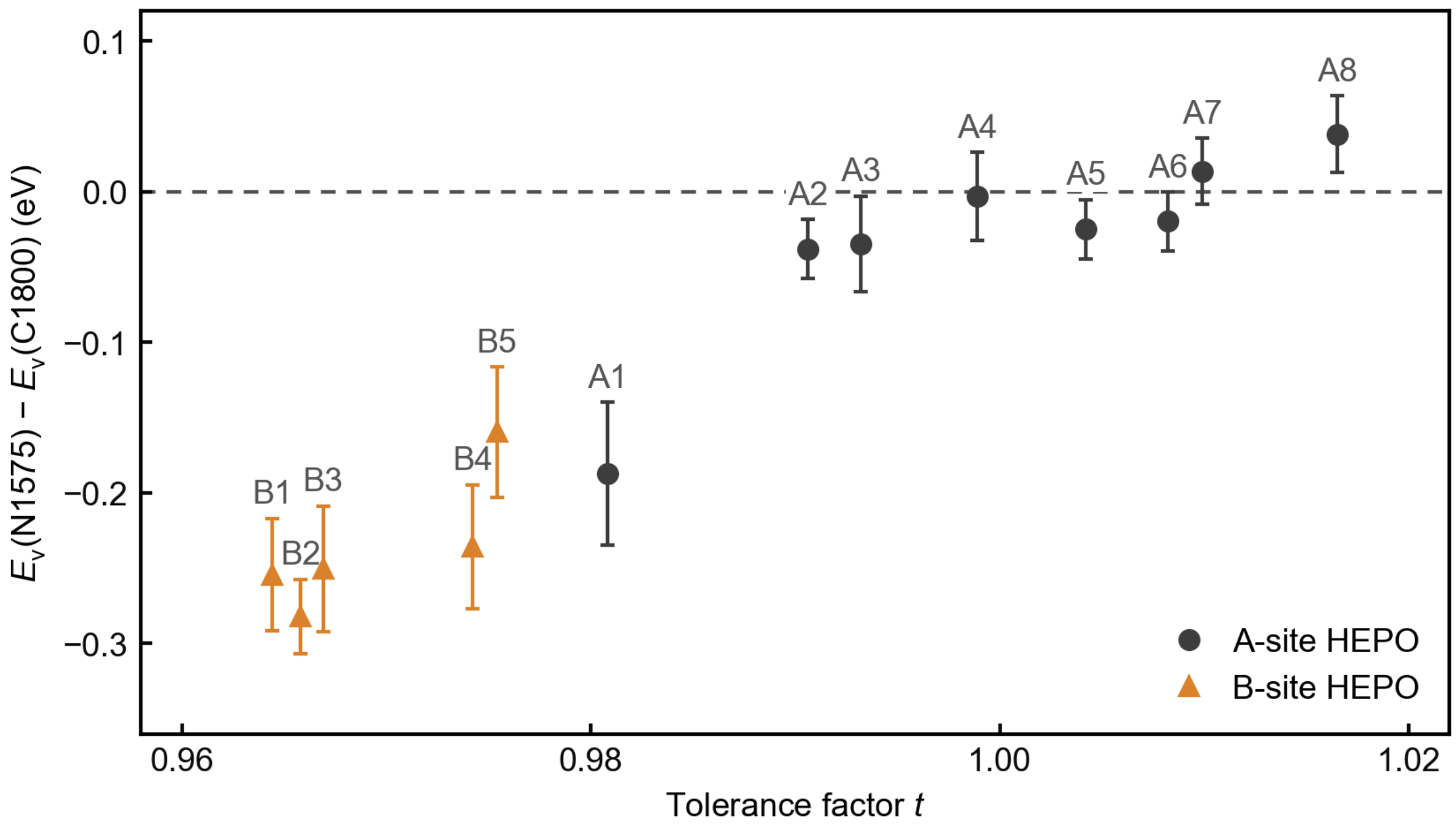

**Fig. S5 Differences in oxygen vacancy formation energy between ~1,700-atom cells.**

The difference $E_v$(N1575) − $E_v$(C1800) between the CHGNet-relaxed commensurate C1800 = 6 × 6 × 10 cell and non-commensurate N1575 = 5 × 7 × 9 cell is plotted against the tolerance factor $t$. For each of the 13 compositions, 20 configurations per cell and 30 sites per configuration were calculated. Error bars indicate the SEs of the differences between the cells, calculated using the 20 configuration-level means in each cell. B2 and B3 have the same $t$ (0.966) and are therefore plotted with a small horizontal offset.